\documentclass[twocolumn,secnumarabic,amssymb, nobibnotes, aps, pra,showpacs,preprintnumbers]{revtex4-1}
\usepackage{amssymb}
\usepackage{array}
\usepackage{graphicx}
\usepackage{amsmath}
\usepackage{amsthm}
\usepackage{amsopn}
\def\uno{\mbox{1 \kern-.59em {\rm l}}}

\def\beq{\begin{equation}}
\def\eeq{\end{equation}}
\def\bea{\begin{eqnarray}}
\def\eea{\end{eqnarray}}

\begin{document}

\title{Generalized Ehrenfest's Equations and phase transition in Black Holes}

\author{Mohammad Bagher Jahani Poshteh$^{1,2}$}
\email{mb.jahani@iasbs.ac.ir}
\author{Behrouz  Mirza$^{1}$}
\email{b.mirza@cc.iut.ac.ir}
\author{Fatemeh  Oboudiat$^{1}$}
\email{f.oboudiat@ph.iut.ac.ir}
\affiliation{$^1$Department of Physics,
Isfahan University of Technology, Isfahan 84156-83111, Iran\\
$^2$Department of Physics, Institute for Advanced Studies in Basic Sciences (IASBS), 45195-1159, Zanjan, Iran}
\begin{abstract}
We  generalize Ehrenfest's equations to systems having two work terms, i.e. systems with three degrees of freedom. For black holes with two work terms  we obtain nine equations instead of two to be satisfied at the critical point of a second order phase transition. We finally generalize this method to a system with an arbitrary number of degrees of freedom and found there is $\frac{N(N+1)^{2}}{2}$ equations to be satisfied at the point of a second order phase transition where $N$ is number of work terms in the first law of thermodynamics.
\end{abstract}


\maketitle

\newpage
\section{Introduction}

Black hole thermodynamics  has become a very important topic over the past few decades \cite{haw}. As a thermodynamic  system, black holes undergo a phase transition shown by Davies for the Kerr-Newman black hole \cite{davies77}. Using the path integral approach, Hawking and Page investigated Schwarzschild-AdS black holes and found the critical temperature and mass at which a phase transition occurs \cite{hp}. This subject has ever since been considered by a number of authors \cite{davies78,davies89,myu,kou,jing,myukim,ann,cao,wang,bis,cai,carlip,cald,cham2,cham1}.\\

It has been shown that the thermal AdS space may collapse to form a black hole at certain temperatures \cite{hp}. This is an example of first order phase transition \cite{cham2} and can also occur for the case of charged and Kerr-Newman black holes in the AdS space \cite{cald,cham1}. The authors of \cite{carlip} argued that  this kind of transition may also occur in flat and de Sitter spaces. They also showed that for black holes below the critical temperature of Hawking-Page phase transition, or the so called 'supercooled black holes', there is a line of first order phase transition in the phase diagram that terminates  at a second order transition point.\\

The aim of this paper is to consider the critical point at which heat capacity diverges. In the laboratory thermodynamics  of the second order phase transition, there is a finite discontinuity in the diagram of specific heat. But as it has been shown by Davies \cite{davies77} that an infinite discontinuity also exist for the critical point in black hole's thermodynamics. So, one can argue that it is not a second order phase transition analogous to the ones in ordinary thermodynamics \cite{soko}. \\

Discontinuity of specific heat has made it  interesting to study the critical phenomena at the divergence point \cite{carlip,cham1,cham2}. Specific heat is approximated around the critical point to a power of temperature, and then the critical exponent $\alpha$ is read of \cite{cham2}. In \cite{cham1} it was shown that for the case of charged black holes, the critical exponent does not depend on the dimensions of the space. Specific heat is also important for the  study of  local stability of black holes \cite{fernando} in the sense that black hole is locally stable if its heat capacity is positive.\\

Here, we use the Clapeyron's and Ehrenfest's scheme to investigate the critical point. In \cite{cham1} Clapeyron equation is used to study the coexistence line of first order phase transition. Ehrenfest's equations must be satisfied at the critical point of a second order phase transition. The critical point is the one in which heat capacity is discontinuous and, as we will see, this is the case for a large class of black holes.\\

 For some years, it was thought that Ehrenfest's equations are not both satisfied at the divergent point of heat capacity; rather numerical investigation was used to show that black holes undergo a glassy phase transition \cite{ban5,ban6}. Later, examination of the divergence point of other parameters that appear in Ehrenfest's equations showed that there was no violation in Ehrenfest's equations, so that the transition is of second order \cite{ban3}. In this paper, we check  Ehrenfest's equations in the point of phase transition in a large class of black holes. Also, we will obtain a generalized form of Ehrenfest's equations for the case of black holes that have more than one work term in their associated first law of thermodynamics and check the relations for phase transition in a Kerr-Newman black hole. It is fascinating that all of the nine generalized Ehrenfest's equations are consistently satisfied at the point of second order phase transition. Finally we generalized Ehrefest equations to the case of arbitrary number of degrees of freedom.\\

This paper is organized as follows. In Section II, we review and check the validity of Ehrenfest's equations for a large number of black holes with two degrees of freedom. In Section III, we generalize Ehrenfest's equations for systems with three degrees of freedom and obtain nine equations to be satisfied at the point of second order phase transition. Then, we check their validity  for Kerr-Newman black hole at the divergence point of specific heat in different ensembles. We generalize our approach to the case of $N$ work terms in Section IV and conclude the paper in Section V.

\section{Second order phase transition for two parameter black holes}
Phase transition in thermodynamic systems is an important area in physics. Ehrenfest categorized phase transitions by using the free energy as a function of thermodynamic variables. In this way, the order of phase transitions is defined as the smallest order of free energy differentials that have noncontinuous behavior through the transition. For example, the transition of solid-liquid-gas in water is a first order phase transition because entropy and volume (which are the first differentials of Gibbs free energy) are noncontinuous through the transition.
In this Section, we investigate the thermodynamic phase transition in a variety of black holes that have just one work term in their associated first law of thermodynamics. Consider the case  of charged black holes. The first law of thermodynamics for this case will be of the following form:
\bea
dM=TdS+\Phi dQ.\label{1law}
\eea
in which, $M$ is the mass of black hole, $T$ is Hawking temperature, $S$ is entropy, and $\Phi$ and $Q$ are electric potential and charge, respectively. The work term in this equation, $\Phi dQ$, which is analogous to $-PdV$ in ordinary thermodynamics, is important for writing Ehrenfest's equations. As we will show soon, Ehrenfest's equations are necessary conditions for the existence of second order phase transitions. To obtain these equations,  we need to develop a suitable form of free energy which is one that ensures the second law of thermodynamics (i.e. increasing the total entropy when it decreases).\\
In canonical ensemble, the suitable form of free energy is defined as follows:
\bea
G=M-TS
\eea
Considering (\ref{1law}), we have:
\bea
dG=-SdT+\Phi dQ.
\eea
We can expand the first differentials $S$ and $\Phi$ in terms of independent parameters $Q$ and $T$:
\bea
dS=(\frac{\partial S}{\partial T})_{Q}dT +(\frac{\partial S}{\partial Q})_{T}dQ=\frac{C_{Q}}{T}dT+ \Phi \alpha dT \label{dss}
\eea
where,  $C_{Q}$ is the specific heat in constant charge and $\alpha=\frac{1}{\Phi}(\frac{\partial S}{\partial Q})_{T}$. At the point of a second order phase transition, we should have continuous $S$ and $\Phi$ through the transition so that $S_{1}=S_{2}$ (where the subscripts $1$ and $2$ are the phases before and after transition). As a result, $dS_{1}=dS_{2}$, and  from (\ref{dss}) we will have:
\bea
-(\frac{\partial Q}{\partial T})_{S}&=&\frac{(C_{Q})_{2}-(C_{Q})_{1}}{\Phi T(\alpha_{2}-\alpha_{1})}\label{e1}
\eea
which is the first Ehrenfest's equation. In the same way, we have for $\Phi$:
\bea
d\Phi=-\Phi \alpha dT+\Phi \kappa dQ
\eea
where, $\kappa=\frac{1}{\Phi}(\frac{\partial \Phi}{\partial Q})_{T}$. From the continuity of $\Phi$ in phase transition, it follows that $d\Phi_{1}=d\Phi_{2}$. So, we find the second Ehrenfest's equation as follows:
\bea
(\frac{\partial Q}{\partial T})_{\Phi}&=&\frac{\alpha_{2}-\alpha_{1}}{\kappa_{2}-\kappa_{1}}.\label{e2}
\eea
For the case of rotating black holes, $Q$ and $\Phi$ are replaced with angular momentum $J$, and angular velocity $\Omega$, respectively.\\
The order of phase transitions can be identified by checking the validity of Ehrenfest's equations at the critical point in which heat capacity diverges. Satisfaction of Ehrenfest's equations has been shown for Reissner-Nordstrom and Kerr black holes in AdS space \cite{ban3,ban1}. We check these equations for Einstein-Gauss-Bonnet ($EGB$), Einstein-Maxwell-Gauss-Bonnet ($EMGB$), Einstein-Yang-Mills-Gauss-Bonnet ($EYMGB$), Kerr, Reissner-Nordstrom ($RN$), Kerr-AdS, RN-AdS, and Horava-Lifshitz ($HL$) with $k=-1$ black holes in the canonical ensemble. \\

Now, let us consider the grand canonical ensemble, in which $J$ or $Q$ are no longer constant and can be exchanged between the  black hole and its environment while their conjugate parameters are  fixed. By choosing the suitable form of free energy, which is $\widetilde{G}=M-TS-\Phi Q$, we can find Ehrenfest's equations for the charged black holes in the grand canonical ensemble:
\bea
-(\frac{\partial \Phi}{\partial T})_{S}&=&\frac{(C_{\Phi})_{2}-(C_{\Phi})_{1}}{TQ(\alpha_{2}-\alpha_{1})}\label{ge1}\\
-(\frac{\partial \Phi}{\partial T})_{Q}&=&\frac{\alpha_{2}-\alpha_{1}}{\kappa_{2}-\kappa_{1}},\label{ge2}
\eea
where, $C_{\Phi}$ is the specific heat at a constant electric potential, $\alpha=\frac{1}{Q}(\frac{\partial S}{\partial \Phi})_{T}$ is analogous to volume expansion, and $\kappa=\frac{1}{Q}(\frac{\partial Q}{\partial \Phi})_{T}$ is analogous to isothermal compressibility. It is straightforward to prove that both Ehrenfest equations (\ref{ge1}) and (\ref{ge2}) are satisfied for $EMGB$, $EYMGB$ and $HL$ black holes.
\section{Generalized Ehrenfest's equations for three parameter black holes}
In this section, we will generalize the Ehrenfest equations to the case of three parameter black holes that have two work terms in their associated first law of thermodynamics. An example of this type of black hole is Kerr-Newman black hole which have both electrical and angular terms in it's mass formula. After a brief notation of thermodynamics of Kerr-Newman black hole we modify Ehrenfest's equations and we obtain nine equations which are supposed to be satisfied at the critical point of second order phase transition. The Smarr mass relation for the Kerr-Newman black hole states that \cite{smarr}:
\bea
M= (2S+\frac{1}{8S}(J^{2}+\frac{1}{4}Q^{4})+\frac{1}{2}Q^{2})^{\frac{1}{2}}.
\eea
 The first law of thermodynamic is
\bea
dM=TdS+\Omega dJ+\Phi dQ \label{dtm}
\eea
Using the first law of thermodynamics one can determine temperature, $T$, angular velocity, $\Omega$, and electrical potential, $\Phi$.

The suitable form of free energy for canonical ensemble is:
\bea
\tilde{G}=M-TS.
\eea
So,  using Equation (\ref{dtm}) we will have:
\bea
d\tilde{G}= -S dT+\Omega dJ+\Phi dQ .\label{dtg}
\eea
We see that $\tilde{G}$ is a function of $T$, $J$, $Q$. We can write $d\tilde{G}$ as follows:
\bea
d\tilde{G}= ( \frac{\partial \tilde{G}}{\partial T})_{J, Q} dT+( \frac{\partial \tilde{G}}{\partial J})_{T, Q} dJ+( \frac{\partial \tilde{G}}{\partial Q})_{T, J}dQ.
\eea
Comparing this equation with  (\ref{dtg}), we obtain
\bea
S= -( \frac{\partial \tilde{G}}{\partial T})_{J, Q}, \quad
\Omega =( \frac{\partial \tilde{G}}{\partial J})_{T, Q}, \quad
\Phi =( \frac{\partial \tilde{G}}{\partial Q})_{T, J}.\quad
\eea
And thereby we obtain  Maxwell's equations:
\bea
( \frac{\partial S}{\partial J})_{T, Q}&=&-( \frac{\partial \Omega}{\partial T})_{J, Q},\nonumber\\
( \frac{\partial S}{\partial Q})_{T, J}&=&-( \frac{\partial \Phi}{\partial T})_{J, Q} , \nonumber\\
( \frac{\partial \Omega}{\partial Q})_{T, J}&=&( \frac{\partial \Phi}{\partial J})_{T, Q}.
\label{max}
\eea
 Ehrenfest equations can be obtained by expanding the  first differentials of free energy ($S$, $\Omega$ and $\Phi$) in terms of the independent parameters ($T$, $J$, and $Q$) and by fixing the differentials at the  transition point. For $S$, we have:
\bea
dS&=&( \frac{\partial S}{\partial T})_{J, Q} dT+( \frac{\partial S}{\partial J})_{T, Q} dJ+( \frac{\partial S}{\partial Q})_{T, J} dQ\nonumber\\
&=& \frac{C_{J, Q}}{T} dT+\Omega \alpha dJ+\Phi \alpha' dQ\qquad
\eea
where,
\bea
\alpha = \frac{1}{\Omega} ( \frac{\partial S}{\partial J})_{T, Q} = - \frac{1}{\Omega} ( \frac{\partial \Omega}{\partial T})_{J, Q},\nonumber\\
\alpha'= \frac{1}{\Phi} ( \frac{\partial S}{\partial Q})_{T, J} = - \frac{1}{\Phi} ( \frac{\partial \Phi}{\partial T})_{J, Q} \nonumber
\eea
Since entropy is continuous, we have $dS_{1}=dS_{2}$, in which the indices show the states before and after the transition.
\bea
& &(C_{J,Q})_{1}+T\Omega \alpha_{1}(\frac{dJ}{dT})_{S}+T\Phi \alpha'_{1}(\frac{dQ}{dT})_{S}\nonumber\\&=&
(C_{J,Q})_{2}+T\Omega \alpha_{2}(\frac{dJ}{dT})_{S}+T\Phi \alpha'_{2}(\frac{dQ}{dT})_{S}\quad \label{ds}
\eea
which can be written in the following form:
\bea
 - ( \frac{dJ}{dT})_{S}= \frac{(C_{J, Q})_{2}-(C_{J, Q})_{1}}{\Omega T(\alpha_{2}-\alpha_{1})} +\frac{\Phi (\alpha'_{2}-\alpha'_{1})}{\Omega (\alpha_{2}-\alpha_{1})} ( \frac{dQ}{dT})_{S}\qquad   \label{i}
\eea
The general expansion of $J$ in terms of $T$, $S$, and $Q$ is:
\bea
dJ=(\frac{\partial J}{\partial T})_{S,Q}dT+(\frac{\partial J}{\partial S})_{T,Q}dS+(\frac{\partial J}{\partial Q})_{S,T}dQ \nonumber
\eea
By, differentiating $J$ with respect to $T$ for a  constant $S$, we have:
\bea
( \frac{dJ}{dT})_{S}= ( \frac{\partial J}{\partial T})_{S, Q}+( \frac{\partial J}{\partial Q})_{S, T} ( \frac{dQ}{dT})_{S}\label{ii}
\eea
Using (\ref{i}) and (\ref{ii}), we find the first two generalized Ehrenfest's equations for systems with three degrees of freedom:
\bea
-( \frac{\partial J}{\partial T})_{S, Q}&=& \frac{(C_{J, Q})_{2}-(C_{J, Q})_{1}}{\Omega T(\alpha_{2}-\alpha_{1})} \label{ee1}\\
-(\frac{\partial J}{\partial Q})_{S, T}&=&\frac{\Phi (\alpha'_{2}-\alpha'_{1})}{\Omega (\alpha_{2}-\alpha_{1})}. \label{ee2}
\eea
Also, from (\ref{ds}), we can write the following relation instead of (\ref{i}):
\bea
-(\frac{dQ}{dT})_{S}= \frac{(C_{J, Q})_{2}-(C_{J, Q})_{1}}{\Phi T(\alpha'_{2}-\alpha'_{1})} +\frac{\Omega (\alpha_{2}-\alpha_{1})}{\Phi (\alpha'_{2}-\alpha'_{1})} ( \frac{dJ}{dT})_{S}\qquad   \label{j}
\eea
It is straightforward to show that:
\bea
(\frac{dQ}{dT})_{S}= ( \frac{\partial Q}{\partial T})_{S, J}+( \frac{\partial Q}{\partial J})_{S, T} ( \frac{dJ}{dT})_{S}\label{jj}
\eea
Comparing (\ref{j}) and (\ref{jj}), we find the third and forth generalized Ehrenfest equations:
\bea
-(\frac{\partial Q}{\partial T})_{S, J}&=&\frac{(C_{J, Q})_{2}-(C_{J, Q})_{1}}{\Phi T(\alpha'_{2}-\alpha'_{1})} \label{ee3}\\
-( \frac{\partial Q}{\partial J})_{S, T}&=&\frac{\Omega (\alpha_{2}-\alpha_{1})}{\Phi (\alpha'_{2}-\alpha'_{1})}\label{ee4}
\eea
Now we write $\Omega$ as a function of $T$, $J$ and $Q$, so we have:
\bea
\label{do}    d\Omega= -\Omega\alpha dT+ \Omega\kappa dJ+ \Phi\chi' dQ.
\eea
Where
\bea
\kappa =\frac{1}{\Omega}(\frac{\partial \Omega}{\partial J})_{T, Q}, \qquad \chi' = \frac{1}{\Phi} (\frac{\partial \Omega}{\partial Q})_{T, J}. \nonumber
\eea
We can repeat the above calculations for $\Omega$. The result is four Ehrenfest equations:
\bea
(\frac{\partial J}{\partial T})_{\Omega, Q}&=&\frac{\alpha_{2}-\alpha_{1}}{\kappa_{2}-\kappa_{1}}\label{ee5}\\
-(\frac{\partial J}{\partial Q})_{\Omega, T}&=&\frac{\Phi(\chi '_{2}-\chi '_{1})}{\Omega(\kappa_{2}-\kappa_{1})}\label{ee6}\\
( \frac{\partial Q}{\partial T})_{\Omega, J}&=&\frac{\Omega(\alpha_{2}-\alpha_{1})}{\Phi(\chi'_{2}-\chi'_{1})}\label{ee7}\\
-( \frac{\partial Q}{\partial J})_{\Omega, T}&=&\frac{\Omega(\kappa_{2}-\kappa_{1})}{\Phi(\chi'_{2}-\chi'_{1})}.\label{ee8}
\eea
For $\Phi$ as a function of $T$, $J$ and $Q$, we have:
\bea
\label{df}     d\Phi= -\Phi\alpha' dT+\Omega\chi dJ+ \Phi\kappa'dQ.
\eea
Where
\bea
\kappa'= \frac{1}{\Phi}(\frac{\partial \Phi}{\partial Q})_{T, J} \qquad
\chi=\frac{1}{\Omega} (\frac{\partial \Phi}{\partial J})_{T, Q}. \nonumber
\eea
And for $\Phi$:
\bea
(\frac{\partial J}{\partial T})_{\Phi, Q}&=&\frac{\Phi(\alpha'_{2}-\alpha'_{1})}{\Omega(\chi_{2}-\chi_{1})}\label{ee9}\\
-(\frac{\partial J}{\partial Q})_{\Phi, T}&=&\frac{\Phi(\kappa'_{2}-\kappa'_{1})}{\Omega(\chi_{2}-\chi_{1})}\label{ee10}\\
(\frac{\partial Q}{\partial T})_{\Phi, J}&=&\frac{\alpha'_{2}-\alpha'_{1}}{\kappa'_{2}-\kappa'_{1}}\label{ee11}\\
-(\frac{\partial Q}{\partial J})_{\Phi, T}&=&\frac{\Omega(\chi_{2}-\chi_{1})}{\Phi(\kappa'_{2}-\kappa'_{1})}.\label{ee312}
\eea
 All the twelve generalized Ehrenfest's equations are presented in Table \ref{eec}. It should be noted that some of the equations are the same and that a total number of nine independent equations exist.

\begin{table*}
\begin{center}
\caption{Generalized Ehrenfest's equations in the canonical ensemble. There are twelve equations which nine of them are independent.}
\vspace{2mm}
\begin{tabular}{|c|c|c|}
\hline

{\bf$S$ fixed }& {\bf$\Omega$ fixed }& {\bf $\Phi$ fixed} \\
\hline
\hline
&&\\
$-(\frac{\partial J}{\partial T})_{S, Q}= \frac{(C_{J, Q})_{2}-(C_{J, Q})_{1}}{\Omega T(\alpha_{2}-\alpha_{1})}$ &
$(\frac{\partial J}{\partial T})_{\Omega, Q}= \frac{\alpha_{2}-\alpha_{1}}{\kappa_{2}-\kappa_{1}}$ &
$(\frac{\partial J}{\partial T})_{\Phi, Q}=\frac{\Phi(\alpha'_{2}-\alpha'_{1})}{\Omega(\chi_{2}-\chi_{1})}$\\
&&\\
$-(\frac{\partial J}{\partial Q})_{S, T}= \frac{\Phi (\alpha'_{2}-\alpha'_{1})}{\Omega (\alpha_{2}-\alpha_{1})}$ &
$-(\frac{\partial J}{\partial Q})_{\Omega, T}= \frac{\Phi(\chi'_{2}-\chi'_{1})}{\Omega(\kappa_{2}-\kappa_{1})}$ &
$-(\frac{\partial J}{\partial Q})_{\Phi, T}=\frac{\Phi(\kappa'_{2}-\kappa'_{1})}{\Omega(\chi_{2}-\chi_{1})}$\\
&&\\
$-(\frac{\partial Q}{\partial J})_{S, T}=\frac{\Omega (\alpha_{2}-\alpha_{1})}{\Phi (\alpha'_{2}-\alpha'_{1})}$&
$-(\frac{\partial Q}{\partial J})_{\Omega, T}=\frac{\Omega(\kappa_{2}-\kappa_{1})}{\Phi(\chi'_{2}-\chi'_{1})}$&
$-(\frac{\partial Q}{\partial J})_{\Phi, T}=\frac{\Omega(\chi_{2}-\chi_{1})}{\Phi(\kappa'_{2}-\kappa'_{1})}$\\
&&\\
$-(\frac{\partial Q}{\partial T})_{S, J}=\frac{(C_{J, Q})_{2}-(C_{J, Q})_{1}}{\Phi T(\alpha'_{2}-\alpha'_{1})}$&
$(\frac{\partial Q}{\partial T})_{\Omega, J}=\frac{\Omega(\alpha_{2}-\alpha_{1})}{\Phi(\chi'_{2}-\chi'_{1})}$&
$(\frac{\partial Q}{\partial T})_{\Phi, J}=\frac{\alpha'_{2}-\alpha'_{1}}{\kappa'_{2}-\kappa'_{1}}$\\
&&\\
\hline
\hline
$\alpha= \frac{1}{\Omega} ( \frac{\partial S}{\partial J})_{T, Q}$ &
$\kappa=\frac{1}{\Omega}(\frac{\partial \Omega}{\partial J})_{T, Q}$ &
$\chi=\frac{1}{\Omega}(\frac{\partial \Phi}{\partial J})_{T, Q}$\\
&&\\
$\alpha'= \frac{1}{\Phi} ( \frac{\partial S}{\partial Q})_{T, J}$ &
$\chi'=\frac{1}{\Phi} (\frac{\partial \Omega}{\partial Q})_{T, J}$ &
$\kappa'= \frac{1}{\Phi}(\frac{\partial \Phi}{\partial Q})_{T, J}$\\
\hline
\end{tabular}
\label{eec}
\end{center}
\end{table*}

\begin{table*}
\begin{center}
{\caption{The singular parameters appearing in the Generalized Ehrenfest's equations for the canonical ensemble.}}

\vspace{5mm}

\begin{tabular}{|c|c|}
\hline $C_{J, Q}$ & $\frac{2S(-4J^{2}-Q^{4}+64S^{2})(4J^{2}+(Q^{2}+8S)^{2})}{A}$ \\
\hline $\alpha$ & $\frac{8\sqrt{8J^{2}S^{3}+2S^{3}(Q^{2}+8S)^{2}}(4J^{2}+(Q^{2}+8S)(Q^{2}+24S))}{A}$ \\
\hline $\alpha'$ & $\frac{8\sqrt{8J^{2}S^{3}+2S^{3}(Q^{2}+8S)^{2}}(4J^{2}(Q^{2}-8S)+(Q^{2}+8S)^{3})}{(Q^{2}+8S)A}$ \\
\hline $\kappa$ & $-\frac{(4J^{2}-(3Q^{2}-8S)(Q^{2}+8S))(4J^{2}+(Q^{2}+8S)^{2})}{JA}$ \\
\hline $\kappa'$ & $\frac{(Q^{2}-8S)(Q^{2}+8S)^{4}+48J^{4}(3Q^{2}+8S)+8J^{2}(Q^{2}+8S)(5Q^{4}+48Q^{2}S+192S^{2})}{Q(Q^{2}+8S)A}$ \\
\hline $\chi$ & $-\frac{8Q(Q^{2}+4S)(4J^{2}+(Q^{2}+8S)^{2})}{A}$ \\
\hline $\chi'$ & $\frac{-16J(Q^{2}+4S)(4J^{2}+(Q^{2}+8S)^{2})}{(Q^{2}+8S)A}$ \\
\hline
\end{tabular}
\bea
A&=&48J^{4}+(3Q^{2}-8S)(Q^{2}+8S)^{3}+8J^{2}(3Q^{4}+32Q^{2}S+192S^{2}). \nonumber
\eea
\label{seec}
\end{center}
\end{table*}

The equations (\ref{ee3}) and (\ref{ee11}) convert to the usual Ehrenfest's equations in the limit of Reissner-Nordestrom black hole. Also, the equations (\ref{ee1}) and (\ref{ee5}) go to the usual Ehrenfest's equations in the limit of the two parametric Kerr black hole.\\
Explicit expressions of singular parameters appearing in the nine generalized Ehrenfest's equations are given in Table \ref{seec}. $A$ is a common factor in the denominator of all the singular parameters. This factor is zero at the critical point. Canceling out this factor is the key to the  satisfaction of Ehrenfest's equations. Now we explain how Equation (\ref{ee1}) is satisfied. Consider the right hand side of Equation (\ref{ee1}). Using the equations for $C_{J, Q}$ and $\alpha$ from Table \ref{seec} we have:
\begin{widetext}
\bea
\frac{(C_{J, Q})_{2}-(C_{J, Q})_{1}}{\Omega T(\alpha_{2}-\alpha_{1})}&=&\frac{2S(-4J^{2}-Q^{4}+64S^{2})(4J^{2}+(Q^{2}+8S)^{2})}{8\sqrt{8J^{2}S^{3}+2S^{3}(Q^{2}+8S)^{2}}(4J^{2}+(Q^{2}+8S)(Q^{2}+24S))} \times\frac{\frac{1}{A_{2}}-\frac{1}{A_{1}}}{\frac{1}{A_{2}}-\frac{1}{A_{1}}} \nonumber\\
&=&\frac{2\sqrt{2}S(4J^{2}S+Q^{4}+16Q^{2}S^{2}+64S^{3})^{\frac{3}{2}}}{J(4J^{2}+Q^{4}+32Q^{2}S+192S^{2})}=-( \frac{\partial J}{\partial T})_{S, Q}
\eea
\end{widetext}
So the first generalized Ehrenfest's equation is satisfied. In the same way, we can show that other equations in Table \ref{eec} are also satisfied.

We can use several ensembles to study the thermodynamics of the black hole. In the grand canonical ensemble, the angular velocity $\Omega$ and the electrical potential $\Phi$ are constants. We also have two mixed ensembles, one is defined by taking $J$ and $\Phi$ to be constant and the other is defined by demanding that $Q$ and $\Omega$ be constant. We called them $J$-fixed and $Q$-fixed ensemble, respectively.
To check the existence of second order phase transition in mixed ensembles we should find generalized Ehrenfest equations in these ensembles by a similar calculations to the previous one for canonical ensemble. But there is a simple method to obtain these relations by comparing energy equations of both ensembles. The suitable form of free energy for $Q$-fixed ensemble is:
\bea
\widetilde{G}=M-TS-J\Omega
\eea
Using the first law of thermodynamics, $dM= T dS+ \Omega dJ+ \Phi dQ$ we have:
\bea
d\tilde{G}= -S dT-Jd\Omega +\Phi dQ
\eea
Comparing this relation and (\ref{dtg}) we see that by replacing \{$J\rightarrow -\Omega$ and $\Omega\rightarrow J$\} or \{$J\rightarrow \Omega$ and $\Omega\rightarrow -J$\} we can obtain differential form of free energy of $Q$-fixed ensemble from the canonical ensemble. We have already obtained the Ehrenfest equations from free energy (table \ref{eec}). So by the above replacements it is possible to obtain a new set of Ehrefest equations in $Q$-fixed ensemble. In a similar way by replacing \{$Q\rightarrow -\Phi$ and $\Phi\rightarrow Q$\} or \{$Q\rightarrow \Phi$ and $\Phi\rightarrow -Q$\} we can obtain another set of generalized Ehrefest equations in $J$-fixed ensemble. By imposing both of the following replacements (\{\{$J\rightarrow -\Omega$, $\Omega\rightarrow J$\}, \{$Q\rightarrow -\Phi$ and $\Phi\rightarrow Q$\}\} or \{\{($J\rightarrow \Omega$, $\Omega\rightarrow -J$\}, \{$Q\rightarrow \Phi$ and $\Phi\rightarrow -Q$\}\}) on canonical ensemble we can find other relations for the grand canonical ensemble. Also by replacing \{$Q\leftrightarrow J$ and $\Phi\leftrightarrow \Omega$\} mixed ensembles convert to each other. So table \ref{eec} can be used to find Ehrenfest equations of all ensembles.
 We can easily demonstrate that the generalized Ehrenfest's equations in the $Q$-fixed and $J$-fixed mixed ensembles are satisfied. But as  shown in \cite{tranah}, the specific heat for constant $\Omega$ and $\Phi$ does not diverge and, thus, so the possibility of a second order transition in the grand canonical ensemble is ruled out.

\section{Higher dimensional generalization of Ehrenfest's equations}
We can generalize the method used in  the previous section to the case of $N$ work terms in the first law of thermodynamics. We assume that we have a free energy with $N$ work terms as below:
\bea
dG=-SdT+\sum_{i=1}^{N}\Omega_{i}dJ_{i}
\eea
To obtain Ehrenfest's equations, we should first write the differentials of the free energy that are $S$ and $\Omega_{i}s$ in terms of independent parameters (i.e., $T$ and $J_{i}s$). So, we have the following  relations:
\bea
dS&=&(\frac{\partial S}{\partial T})_{\{J_{i}\}}dT+\sum^{N}_{j=1}(\frac{\partial S}{\partial J_{j}})_{T,\{J_{i\neq j}\}}dJ_{j}\nonumber\\
d\Omega_{i}&=&(\frac{\partial \Omega_{i}}{\partial T})_{\{J_{i}\}}dT+\sum^{N}_{j=1}(\frac{\partial \Omega_{i}}{\partial J_{j}})_{T,\{J_{i\neq j}\}}dJ_{j}\label{tt}\label{3}
\eea
We have $N+1$ equations similar  to (\ref{3}), differentiating any of which with respect to $T$ leads to Ehrenfest's equations. We can choose any one of $N$ work terms in Equations (\ref{tt}) to write the following relations:
\bea
\left(\frac{dJ_{i}}{dT}\right)_{\{\Omega_{k}\}}=\alpha_{i}+\sum_{j\neq i}\beta_{ij}\left(\frac{dJ_{j}}{dT}\right)_{\{\Omega_{k}\}}\label{4}
\eea
where, $\alpha_{i}$ and $\beta_{ij}$ are some constants. Each set of  $N+1$ equations of (\ref{tt}) generates $N$ equations similar to  (\ref{4})  each of which  leads to $N$ Ehrenfest's equations. So far we have $N^{2}(N+1)$ equations  some of which  are the same. Sentences like $\left(\frac{\partial J_{i}}{\partial J_{j}}\right)_{T,\{\Omega_{k}\}}$ and $\left(\frac{\partial J_{i}}{\partial J_{j}}\right)_{T,S}$ lead to the same Ehrenfest's equations under exchange of $i\leftrightarrow j$. We have $N^{2}(N-1)$ terms like $\left(\frac{\partial J_{i}}{\partial J_{j}}\right)_{T,\{\Omega_{k}\}}$ ($i,k=1,2,\ldots ,N\qquad i\neq j$) and $N(N-1)$ terms like $\left(\frac{\partial J_{i}}{\partial J_{j}}\right)_{T,S}$. So, the final number of Ehrenfest's equations is given by:
\bea
& &N^{2}(N+1)-\frac{1}{2}\left(N^{2}(N-1)+N(N+1)\right)\nonumber\\&=&\frac{N(N+1)^{2}}{2}\label{number}
\eea
It is clear that we have two equations for $N=1$, nine equations for $N=2$, etc.
\\Also, we can exploit the same argument to determine  the number of generalized Ehrenfest's equations for Myers-Perry black holes \cite{myers} that are the generalized form of  Kerr-Ads black holes in higher dimensions, with rotation in more than one plane. In higher dimensions, the number of rotating planes will be as follows:
\bea
N\leq \left[\frac{d-1}{2}\right]\nonumber
\eea
where, $d$ is the number of space-time dimensions and brackets give the integer part. For example, we have two rotating planes for a five dimensional space-time and  three for a seven dimensional one. We  have nine generalized Ehrenfest's equations for a five dimensional Myers-Perry (MP) black hole and 24 equations for a seven dimensional MP black hole.

\section{Conclusions}
In this paper, we considered the thermodynamic phase transition in black holes strongly focusing  on Ehrenfest's equations. For a large class of black holes which have just one work term in their first law of thermodynamics, we showed that at a second order phase transition in the canonical and grand canonical ensemble the Ehrenfest equations are satisfied.
 
 We also generalized Ehrenfest's equation to the case of systems with three degrees of freedom. We  obtained nine Ehrenfest's equations that are satisfied at the critical points. As an example, we considered the  Kerr-Newman black hole in the canonical ensemble. We showed that the generalized Ehrenfest's equations are satisfied at the critical point in which heat capacity diverges. So, the infinite discontinuity of specific heat in black hole thermodynamics causes no problem for a second order phase transition to occur. In the case of the Kerr-Newman black hole, we found two mixed ensembles that have their own generalized Ehrenfest's equations. Finally, we generalized this approach to the case of $N$ work terms in the first law of thermodynamic  which can be used to check the order of  transitions in  black holes that have more than two work terms such as Myers-Perry black hole in higher dimensions.


\end{document}